%% file: class.tex
\documentclass[sigconf, 11pt, letter]{acmart}

\renewcommand\footnotetextcopyrightpermission[1]{} 
\usepackage{courier}
\usepackage{amsmath,amssymb,amsfonts}
\usepackage{algorithmic}
\usepackage{graphicx}
\usepackage{enumitem}
\usepackage{url}
\usepackage{hyperref}
\usepackage{comment}

\usepackage{float}
\usepackage[export]{adjustbox}
\usepackage{xcolor}

\def\BibTeX{{\rm B\kern-.05em{\sc i\kern-.025em b}\kern-.08em
    T\kern-.1667em\lower.7ex\hbox{E}\kern-.125emX}}

\usepackage{booktabs} 
\usepackage{paralist}

\usepackage[ruled]{algorithm2e}

\begin{document}

\title{Quality of Service in IEEE 802.11 WLANs: An Experimental Study} 
\author{Farideh Parastar*, Shian J. Wang}

\affiliation {*Ferdowsi University of Mashhad}

\begin{abstract}
While the IEEE 802.11 protocol is being widely used, it is not specifically designed to handle multimedia traffic, which covers an important portion of the Internet traffic today. Voice and video multicast streaming is inefficient, as there is lack of transmission reliability in such approaches where delay is not guaranteed. In this work, we  focus on the performance of Enhanced Distributed Channel Access (EDCA) mechanism proposed by IEEE 802.11e which provides traffic prioritization and since most of the previous works are done based on simulation results, we test the performance of this protocol in a real platform using sofware Defined Networks. We then validate the impact of different EDCA parameters ( e.g., AIFS and TXOP) by tuning them to see their impact on the delay of the network and eventually on network traffic.
\end{abstract}

\maketitle
\input{textsum}

\end{document}

%% file: textsum.tex
\section{Introduction}
Today IEEE 802.11 WLAN standard is being accepted and widely used  for many environments  \cite{IEEE802.11b}. Deployment of  the WLAN is a cheap and easy process. On the other hand, real-time traffic is becoming significant part of the internet.  Unlike best effort traffic,  real-time traffic needs Quality of Service (QoS) support. Providing QoS support is a challenging task in IEEE 802.11 WLAN since there is no method for differentiating  traffic.  IEEE 802.11e MAC protocol ~\cite{IEEEP802.11ED} provides QoS support for different types of traffic by adjusting the MAC protocol parameters. There is another upcoming amendment namely IEEE 802.11aa ~\cite{802.11a}. The goal of the mechanisms proposed in 802.11aa is to address multimedia streaming issues in 802.11 networks.  802.11e amendment has been studied under many analytical and experimental studies, but to our knowledge, most of these studies are based on simulation results and there are less works which test the performance of this amendment in a real testbed. In this paper, we aim to test and evaluate this protocol based on parameters proposed in previous studies simulations and compare the results to the existing Distributed Coordination Function (DCF) and observe the performance of Enhanced Distributed Channel Access (EDCA)  in real testbed rather than simulations. \\
The rest of this paper is organized as follows: In section \ref{overview}, we give a summary about the channel access mechanisms of 802.11 WLAN and the 802.11e amendment. Section \ref{related} is about the related work which we have extracted from the literature. In section \ref{implementation}, we describe the configuration of platform we used. In Section \ref{result} we illustrate the QoS limitations of DCF and we demonstrate the performance of EDCA through experiments. In the last part we talk about EDCA parameters impact on mean delay. Finally, we conclude the paper. 

\section{Overview of IEEE 802.11e Channel Access} \label{overview}
\subsection{802.11  Distributed Coordination Function (DCF)}
\begin{figure}[h]
\centering
\includegraphics[width=8cm, height=5cm]{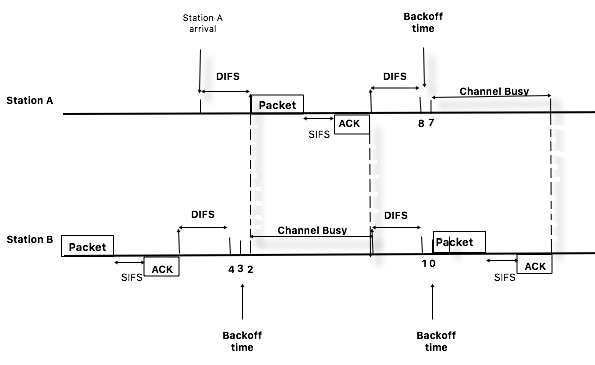}
\caption{ This figure illustrate the basic access mechanism in 802.11 DCF. Station B has already sent a packet successfully and wants to send the second one. In between station A arrives, and finds the channel idle for duration of DIFS. Thus, it initiates the transmission and at the same time station B sense the channel busy, thus it halts backoff counter until it sense the medium idle again.\label{fig:DCF}}
\end{figure}

DCF is an access coordination function defined in 802.11 standard, which works based on carrier sense multiple access collision avoidance (CSMA/CA) mechanism. Two stations sensing the channel idle at the same time may end up with collision. As a part of collision avoidance mechanism, after sensing the channel idle, each station waits for an extra Distributed Interframe Space (DIFS) duration.  Only if the channel remains idle for DIFS time period, the station is allowed to initiate the transmission otherwise the transmission is differed. The random time duration is specified as a multiple of a slot time ~\cite{Mangold}. If the station senses  the channel busy in DIFS duration, it has to persist monitoring until it senses the medium idle for another DIFS duration. After this duration, station waits for  an additional random backoff time which is between [0, CW-1] where CW is Contention Window and depends on the number of retransmitted packets. CW starts from CWmin value and after each collision it grows exponentially until it reaches CWmax value. If station finds the medium busy during backoff time, it freezes the counter and keeps monitoring the channel to find it idle for DIFS duration.  Then it resumes the frozen random backoff time to count down. The random backoff is decremented by one after each idle slot time on medium. A successful transmission is followed by an immediate acknowledgment, since stations cannot find collisions by listening to their own transmission. If a transmission fails, the retransmission will be initiated with a doubled size of contention window for clashed stations to reduce the probability of collisions. If the number of retransmission reaches a limit, the packet will be discarded. Stations also need to wait an extra post backoff time after each successful transmission. Figure \ref{fig:DCF} illustrates the basic access mechanism of DCF. \\
To solve the problem of hidden stations, there is another optional mechanism called Request To Send/ Clear To Send.  The RTS and CTS frames include the information of how long it takes to transmit the next data frame. A station that wishes to transmit data, after detecting medium idle for a DIFS duration followed by a random backoff, sends a short RTS frame, and the receiver of this frame, will send a CTS frame after SIFS duration (which is 10 $\mu$s for IEEE 802.11g). In this way, stations either close to the transmitting or receiving stations will hear these frames and stop transmitting data for a duration mentioned in RTS/CTS frames. More specifically, their timer called Network Allocation Vector, NAV, is set ~\cite{Mangold} to a time duration mentioned in these packets. RTS/CTS mechanism is good when there are frames with longer size to transmit. By this mechanism a long frame can be transmitted at once, and fragmentation is no more needed.

\subsection{802.11e Enhanced Distributed Channel Access (EDCA)}
802.11 e is an amendment to the IEEE 802.11 Wireless LANs (WLANs) standard, which provides Quality of Service (QoS) support at the MAC layer. To support QoS, two mechanisms have been proposed: Enhanced Distributed Channel Access (EDCA) and  Hybrid Coordination Function (HCF). Our main focus is on the former mechanism. \\
EDCA is improvement of DCF in 802.11 legacy. DCF does not provide any prioritization for different traffic types. Instead,  EDCA provides traffic differentiation and defines four Access Categories (AC). In EDCA mechanism, smaller CWs are assigned to ACs with higher priority. CWs are initialized uniquely for each AC. After sensing channel idle, each AC within a station, starts a backoff time for a period of Arbitration Interframe Space (AIFS). AIFS is calculated as follows:
\begin{equation}
AIFS=SIFS + AIFSN * a SlotTime
\end{equation}

Where AIFSN is  AIFS number determined by AC. \\
When medium is idle for AIFS period, each AC sets a counter, which is a random number between [0, CW[AC]] \cite{Mangold}, and decrements it by one. If an AC finds the medium busy, before the counter reaches zero, it should wait for another AIFS period. Each AC behaves like a station. When a collision happens between ACs within the same station, the AC with higher priority is able to successfully transmit while AC with lower priority will suffer from virtual collision. After any unsuccessful transmission, a new CW is set to reduce the collision probability. This CW is set according to the following formula:
\begin{equation}
new CW [AC] >= ((Old CW[AC]+1)* 2)-1
\end{equation}
IEEE 802.11e also defines a transmission opportunity (TXOP) limit as the interval of time in which a station is  allowed to transmit multiple data frames from the same AC with a SIFS gap between an ACK and the data frame \cite{Mangold,MAC,Mangold2,kong}.

\section{Related Work} \label{related}
EDCA performance for QoS has been explored extensively in the literature. The authors of \cite{edca} and \cite{edca2} have illustrated limitations of QoS in 802.11, and they evaluate the performance of 802.11e through some simulations. In \cite{optimal} they have implemented an algorithm which calculates the optimal values of EDCA parameters. This algorithm is based on mathematical analysis of throughput and delay. In \cite{experimental} they have studied two applications of EDCA: traffic engineering and service guarantee and they showed traffic engineering is well supported by EDCA when there is only UDP traffic in WLAN~\cite{experimental}. Some other works focused on QoS in terms of multimedia streams. In \cite{video1} and \cite{video2}, a QoS framework has been proposed that maps categorized video packets onto the relative differentiated service provided by the wireless channel using a predetermined pricing model. In \cite{parameter} they have studied the impact of TXOP, CWmin and AIFS on the throughput, and compared their results to the proposed analytical models. 

Machine Learning techniques are used in many network sub-fields including resource allocation, parameter optimization, traffic prediction \cite{prediction2} and classification \cite{classification}, congestion control \cite{congestion}. Recently Reinforcement learning has gain attention by the research community for communication protocol design. Authors in \cite {pasandi2019challenges} ~\cite{deeplearningmac} use machine learning to overcome DCF issues as hidden terminal problems, or dealing with high number of nodes by optimizing the channel by different MAC functionalities. In their approach a MAC protocol is decomposed l into a set of modular blocks. In different network scenarios, different blocks (e.g., contention window, backoff) are selected by the reinforcement learning agent. Authors in \cite{deep} tune the contention window parameter of backoff mechanism. Their learning-based approach shows that resetting contention window to zero in in fact harmful in there is a dense number of users contending for the meduim. Authors in \cite {pasandi2019towards} and \cite{poster} describe a broader overview of a framework that shows how ML techniques are leveraged for centralized and distributed ML agents to design MAC protocols. 
\section{Implementation} \label{implementation}
\subsection{Platform}
We have used our wireless testbed which is based on RaspberryPi B/B+ model, Power-over-Ethernet (PoE) switches, and a standard PC for the server. For our tests, we added a WiFi dongle on each node and deployed image on each  RaspberryPi to control nodes and monitor the results.

\section{Results} \label{result} 
\subsection{DCF Performance}
We have proposed a scenario namely scenario 1 which is roughly based on a simulation in \cite{edca} but with less number of nodes.  In this scenario, each station  sends three types of traffic. In our tests 802.11g is set as PHY layer with sending rate up to 54 \textit{Mb/s}. All the tests in this section are repeated for 10 times, and we noticed that the results are stable and there is no significant variant in data. \\
 To test the DCF QoS limitations, we run scenario 1. All the stations operate in infrastructure mode meaning they send and receive data from Access Point (AP).  Table \ref{my-label} shows the parameters used for different traffics. At the beginning there are only 2 stations contending for the channel. Later we increase the load by incrementing the number of stations up to 6. Table \ref{DCF-table} shows mean delays of each type of traffic against the number of stations.  As we notice, there is no important traffic divergence in DCF, meaning all stations have the same right way to get the channel. When the load increases, the mean delay of voice grows from 1.25 \textit{ms} up to almost 51.447 \textit{ms} which is almost similar to video and BE latency. Generally,  when the number of stations increases, the average delay for all kinds of traffic increase in the same manner which can cause QoS problems for real-time streams.
\begin{table}[H]
\centering
\caption{DCF Test Parameters}
\label{my-label}
\setlength\tabcolsep{2.5pt}
\begin{tabular}{|c|c|c|c|}
\hline
                   & Voice     & Video      & Best Effort \\ \hline
Transport Protocol & UDP       & UDP        & UDP         \\ \hline
Packet Size        & 160 Bytes & 1280 Bytes & 1500 Bytes  \\ \hline
Sending Rate       & 64 kb/s   & 640 kb/s   & 960 kb/s    \\ \hline
\end{tabular}
\end{table}

\begin{table}[H]
\centering
\caption{Mean Delay With DCF Mechanism}
\label{DCF-table}
\begin{tabular}{@{}llll@{}}
\toprule
\begin{tabular}[c]{@{}l@{}}Number\\ of STAs\end{tabular} & Voice (ms)                            & \multicolumn{1}{c}{Video (ms) }       & \multicolumn{1}{c}{BE (ms) }          \\ \midrule

{ 2}                                 & { 1.25}  & { 1.51}     & { 1.91}   \\
                                                         &                                  &                                 &                                 \\
 
{ 3}                                 & { 10.53}  & { 10.96}    & { 11.41}  \\
                                                         &                                  &                                 &                                 \\
 
{ 4}                                 & { 21.01} & { 21.47}     & { 22.10}     \\
                                                         &                                  &                                 &                                 \\
 
{ 5}                                 & { 33.62} & { 32.94}  & { 32.68}  \\
                                                         &                                  &                                 &                                 \\

{ 6}                                 & { 51.44}  & {49.11} & { 49.66} \\ \bottomrule
\end{tabular}
\end{table}

\subsection{EDCA Performance}
To evaluate the performance of EDCA, we have tested the same scenario as before  with the set of default parameters defined by the standard. List of parameters used for the test are summarized in Table \ref{my-label2}.  EDCA provides traffic differentiation by giving higher priority to real-time streams.  The mean delay against number of stations is shown in table \ref{EDCAtest} . When the load increases, the mean delays for both voice and video traffic remain small, while with DCF mechanism, average delay for all three types of traffic do not have significant variation. This test demonstrates that EDCA can provide the expected traffic differentiation. By adding the sixth station the delay for both voice and video grows drastically. This is due to small CW values that have been defined by the standard. When the number of clients becomes larger, the probability of having the same CW value for 2 clients at the same time also increases and it causes collisions followed by packet drops.  In general, the higher priority flow, has lower mean latency, and mean delay of BE has remarkable variation compared to multimedia traffics. 

\begin{table}[H]
\centering
\caption{EDCA Test Parameters}
\label{my-label2}
\setlength\tabcolsep{5pt}
\begin{tabular}{|c|c|c|c|}
\hline
                   & Voice & Video & Best Effort \\ \hline
Transport Protocol & UDP   & UDP   & UDP         \\ \hline
CWmin              & 7     & 15    & 31          \\ \hline
CWmax              & 15    & 31    & 1023        \\ \hline
AIFS               & 2     & 2     & 7           \\ \hline
\end{tabular}
\end{table}
\begin{table}[H]
\centering
\caption{Mean Delay With EDCA }
\label{EDCAtest}
\begin{tabular}{@{}llll@{}}
\toprule
\begin{tabular}[c]{@{}l@{}}Number\\ of STAs\end{tabular} & Voice (ms)                           & \multicolumn{1}{c}{Video (ms) }     & \multicolumn{1}{c}{BE (ms) }         \\ \midrule

{ 2}                                 & { 0.88}   & { 1.11} & { 2.20}  \\
                                                         &                                  &                               &                                \\
 
{ 3}                                 & { 2.26}      & { 2.62}   & { 7.03}  \\
                                                         &                                  &                               &                                \\
 
{ 4}                                 & { 7.52}      & { 7.81}   & { 15.85} \\
                                                         &                                  &                               &                                \\
 
{ 5}                                 & { 7.49}      & { 7.92} & { 15.22} \\
                                                         &                                  &                               &                                \\
 
{ 6}                                 & { 274.43} & { 283.56} & { 386.73}  \\ \bottomrule
\end{tabular}
\end{table}

\subsection{DCF and EDCA Performance under Saturated Network}
\begin{figure}
\centering
\includegraphics[width=8cm, height=7cm]{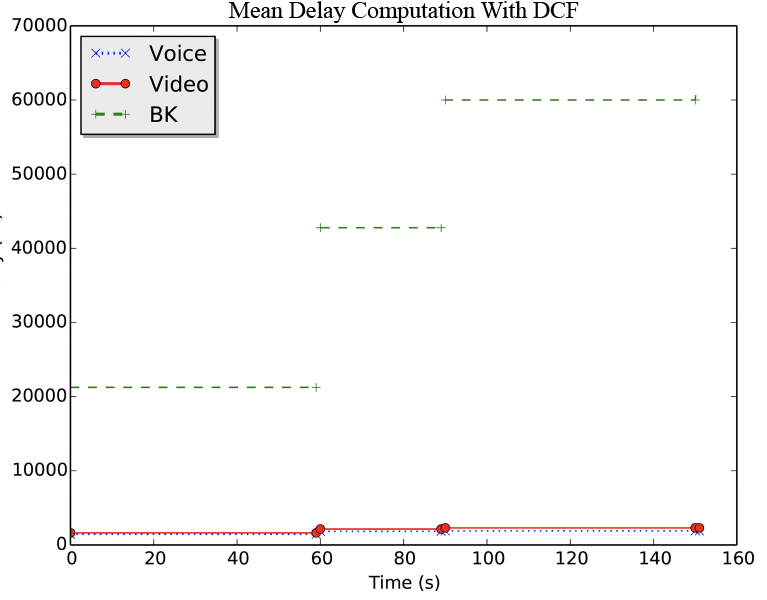}
\caption{Average Delay for Different Types of Traffics With DCF Under Saturated Network  \label{fig:DCFbest}}
\end{figure}

We run set of tests in which we want to evaluate the performance of the DCF and EDCA when the network is saturated with low priority BK  flows. At the start point, there are 3 stations that send voice, video and BK traffics, respectively. The sending rate and packet size for voice and video traffics are specified in Table \ref{my-label}, and BK station sends as many packets as possible to pressurize the network. After 60 seconds, another station with BK traffic is added, and after 90 seconds we add 2 more stations with BK traffic. The whole test lasts for 150 seconds. The results shown in Figure \ref{fig:DCFbest} demonstrate that adding low priority traffic will not have great impact on performance of higher priority streams. This result matches the work done in the literature where it shows that when we keep the packet rate of a host under a limiting value, the host benefits from low delays with any control mechanism. Thus, the performance of hosts with low packet rate is not affected irrespective of the greedy behavior of other stations, and short packet rate can guarantee QoS for time-sensitive flows. We run the same test with EDCA and  the results show that saturating the network by adding low priority flows cannot variate the delay for voice and video streams, and it remains almost stable during the test period.

%\usepackage{booktabs}

\begin{figure}
\centering
\includegraphics[width=8cm, height=7cm]{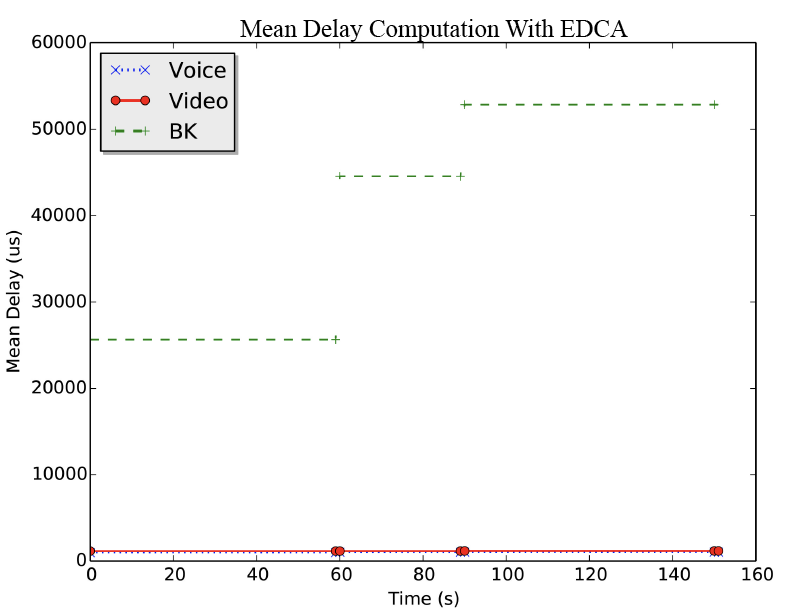}
\caption{Average Delay for Different Types of Traffics With EDCA Under Saturated Network  \label{fig:DCF1width}}
\end{figure}

\subsection{Validation Of EDCA Parameters}
In 802.11e each kind of traffic is directed to a specific queue which is assigned with different MAC layer parameters. In set of experiments we want to clarify the impression of TXOP and AIFS on mean delay under saturated network where each station has at least one packet to transmit. In scenario 1, there are 2 hosts contending to transmit their packets. One is sending video and  the other is sending BK traffic, and each test lasts for 60 seconds. \\
 The effect of TXOP is straightforward. When TXOP value becomes larger, the  throughput for that station gets larger, while the mean delay will be decreased as it can send multiple data . On the other hand,  the average delay for other stations will be increased since they have to wait for a longer time to transmit their packets. \\ Table \ref{txop} demonstrates the effect of TXOP on mean delay of video and BK flows.\\
 
A station needs to sense the channel idle for time interval of AIFS to be able to resume its backoff counter. When the medium becomes busy the MAC backoff counter halts, and it will be resumed when the channel is  free for duration of AIFS. If AIFS is increased for a station, the transmission opportunities will be reduced, since it needs to wait for longer time for packet transmission. When the network is lightly loaded, the AIFS will have minor impact on the delay, but when the load increases, stations with larger AIFS, will be punished.\\
Table \ref{aifs} shows the effect of increasing AIFS on Video traffic delay under saturated network.
%\begin{figure}
%\centering
%\includegraphics[width=7cm, height=6cm]{3txopdot.pdf}
%\caption{This figure illustrate that by increasing TXOP of video, the mean delay for this flow will be reduced, while this value for BK traffic will be increased.    \label{txop}}
%\end{figure}

\begin{table}[h]
\centering
\begin{tabular}{|c|c|c|}

\hline
\textbf{TXOP Period (us)} & \textbf{BK} & \textbf{Video} \\ \hline
10                        & 1.1         & 60             \\ \hline
100                       & 0.9         & 120            \\ \hline
150                       & 0.05        & 140            \\ \hline
\end{tabular}
\caption{This figure illustrate that by increasing TXOP of video, the mean delay for this flow will be reduced, while this value for BK traffic will be increased. \label{txop}}
\end{table}

\begin{table}
\centering
\caption{AIFS impact On Mean Delay of Video Traffic}
\label{aifs}
\begin{tabular}{|l|l|}
\hline
AIFS  & Mean Delay (ms) \\ \hline
3    & 1.43         \\ \hline
7    & 1.52         \\ \hline
12   & 3.75         \\ \hline
14   & 15.89        \\ \hline
\end{tabular}
\end{table}

\section{Conclusion}
In this research, QoS limitations by DCF mechanism are discussed. We have demonstrated that EDCA supports service differentiation between different types of flow with respect to various parameters. We have shown CW values proposed by  802.11e amendment are not optimal when the number of clients is large.  The evaluation of results also shows that limited packet sending rate guarantees low delays for traffics even in saturated case.
\bibliographystyle{ACM-Reference-Format}
\bibliography{ref}